\documentstyle[12pt,epsfig]{article}
\advance\textheight by \topskip
\begin{document}
\tolerance=100000
\thispagestyle{empty}
\setcounter{page}{0}

\newcommand{\be}{\begin{equation}}
\newcommand{\ee}{\end{equation}}
\newcommand{\br}{\begin{eqnarray}}
\newcommand{\er}{\end{eqnarray}}
\newcommand{\ba}{\begin{array}}
\newcommand{\ea}{\end{array}}
\newcommand{\bi}{\begin{itemize}}
\newcommand{\ei}{\end{itemize}}
\newcommand{\bn}{\begin{enumerate}}
\newcommand{\en}{\end{enumerate}}
\newcommand{\bc}{\begin{center}}
\newcommand{\ec}{\end{center}}
\newcommand{\ul}{\underline}
\newcommand{\ol}{\overline}
\newcommand{\ar}{\rightarrow}
\newcommand{\sm}{${\cal {SM}}$}
\newcommand{\susy}{{{SUSY}}}
\newcommand{\Dir}{\kern -6.4pt\Big{/}}
\newcommand{\Dirin}{\kern -10.4pt\Big{/}\kern 4.4pt}
\newcommand{\DDir}{\kern -10.6pt\Big{/}}
\newcommand{\DGir}{\kern -6.0pt\Big{/}}
\def\gluino{\ifmmode{\mathaccent"7E g}\else{$\mathaccent"7E g$}\fi}
\def\photino{\ifmmode{\mathaccent"7E \gamma}\else{$\mathaccent"7E \gamma$}\fi}
\def\mgluino{\ifmmode{m_{\mathaccent"7E g}}
	     \else{$m_{\mathaccent"7E g}$}\fi}
\def\taugluino{\ifmmode{\tau_{\mathaccent"7E g}}
	     \else{$\tau_{\mathaccent"7E g}$}\fi}
\def\mphotino{\ifmmode{m_{\mathaccent"7E \gamma}}
	     \else{$m_{\mathaccent"7E \gamma}$}\fi}
\def\ML{\ifmmode{{\mathaccent"7E M}_L}
	     \else{${\mathaccent"7E M}_L$}\fi}
\def\MR{\ifmmode{{\mathaccent"7E M}_R}
	     \else{${\mathaccent"7E M}_R$}\fi}

\def\Ord{\buildrel{\scriptscriptstyle <}\over{\scriptscriptstyle\sim}}
\def\OOrd{\buildrel{\scriptscriptstyle >}\over{\scriptscriptstyle\sim}}
\def\pl #1 #2 #3 {{\it Phys.~Lett.} {\bf#1} (#2) #3}
\def\np #1 #2 #3 {{\it Nucl.~Phys.} {\bf#1} (#2) #3}
\def\zp #1 #2 #3 {{\it Z.~Phys.} {\bf#1} (#2) #3}
\def\pr #1 #2 #3 {{\it Phys.~Rev.} {\bf#1} (#2) #3}
\def\prep #1 #2 #3 {{\it Phys.~Rep.} {\bf#1} (#2) #3}
\def\prl #1 #2 #3 {{\it Phys.~Rev.~Lett.} {\bf#1} (#2) #3}
\def\mpl #1 #2 #3 {{\it Mod.~Phys.~Lett.} {\bf#1} (#2) #3}
\def\rmp #1 #2 #3 {{\it Rev. Mod. Phys.} {\bf#1} (#2) #3}
\def\sjnp #1 #2 #3 {{\it Sov. J. Nucl. Phys.} {\bf#1} (#2) #3}
\def\cpc #1 #2 #3 {{\it Comp. Phys. Comm.} {\bf#1} (#2) #3}
\def\xx #1 #2 #3 {{\bf#1}, (#2) #3}
\def\preprint{{\it preprint}}

\begin{flushright}
{Cavendish-HEP-96/16}\\
{DFTT 52/96}\\ 
{UGR-FT-70}\\
{August 1996\hspace*{.5 truecm}}\\
\end{flushright}

\vspace*{\fill}

\begin{center}
{\Large \bf Light, long-lived and secluded:\\ 
can gluinos be driven out from LEP1 data ?}\\[0.5 cm]
{\large S. Moretti$^{a,b}$, R. Mu\~noz-Tapia$^c$ and 
	   K. Odagiri$^{a}$}\\[0.4 cm]
{\it a) Cavendish Laboratory, University of Cambridge,}\\
{\it Madingley Road, Cambridge CB3 0HE, UK.}\\[0.25cm]
{\it b) Dipartimento di Fisica Teorica, Universit\`a di Torino,}\\
{\it and I.N.F.N., Sezione di Torino,}\\
{\it Via Pietro Giuria 1, 10125 Torino, Italy.}\\[0.25cm]
{\it c) Dpto F\'{\i}sica Te\'orica y del Cosmos,}\\
{\it Universidad de Granada, Granada 18071, Spain.}\\[0.5cm]
\end{center}
\vspace*{\fill}

\begin{abstract}
{\noindent 
\small 
We briefly report about a possible settlement of
the still ongoing dispute concerning the existence 
of SUSY signals in 4jet events at LEP1.
We base our arguments on a simple selection strategy
exploiting secondary vertex tagging and kinematical constraints, 
which could allow
one to access or exclude
gluino events for a broad range of masses and lifetimes.
}
\end{abstract}
\vskip1.0cm
\hrule
\vskip0.25cm
\noindent
E-mails: moretti,odagiri@hep.phy.cam.ac.uk, rmt@ugr.es.
\vspace*{\fill}
\newpage
\subsection*{Introduction}
\noindent
If the results of the LEP1 measurements were pieces of a jigsaw puzzle
reproducing the edifice of the Standard Model, 
then one could well question
that some of these are apparently not so perfectly shaped, i.e.,
they fit into their original location only with some effort.
This is certainly the case for the determination of $\Gamma_c$ and 
$\Gamma_b$, the partial widths of the $Z$ into $c$ and $b$ quarks,
for which claims of Supersymmetry (SUSY) hints have been made \cite{Rs}.
Indeed, there is another controversy still open along the same lines. 
Somewhat less glamorous but not for this less important 
is the possibility of gluino events being present in 4-jet decays 
\cite{colour,OPAL}.

The story goes as follows. The colour factors $C_A$, $C_F$ and $T_F$ 
of QCD \cite{Quigg} can be measured by fitting some angular 
distributions\footnote{That
is, the angles of Bengtsson-Zerwas, of
K\"orner-Schierholz-Willrodt, of
Nachtmann-Reiter and that between the two least energetic jets
\cite{angles}.} whose shape
significantly depends on the partonic composition of the $\alpha_s^2$
decays of the $Z$. Then one obtains that,  
although the experimental measurements agree well with ordinary 
QCD, it is not possible to rule out its Supersymmetric
version, which predicts that light gluinos $\gluino$ 
(the SUSY partners of the gluons) can be produced at LEP1 energies 
\cite{OPAL}.
In detail, gluinos with a mass $\mgluino\OOrd2$ GeV yield an expectation 
value for $T_F/C_F$ that is 
within one standard deviation of the measured one \cite{OPAL}. 
In fact, if the gluino mass is light enough \cite{Masiero-Farrar-Quarks94}, 
such particles should
be produced in the process $e^+e^-\ar Q\bar Q
\gluino\gluino$ via a $g^*\ar \gluino\gluino$
splitting \cite{window,epem}. Since
gluinos are coloured fermions, such events would enter into the sample
with a behaviour similar to that of $Q\bar Qq\bar q$ events.
Naively, one could well say that the
total number of flavours $N_F$ of the theory is apparently
increased, such that, a
\susy\ signal reveals itself as an enhancement of
$T_R\equiv N_FT_F$. Such kind of new particles are at present 
still compatible with the experiments \cite{Farrar-review,Kileng-Osland1}.

The reason why experiments have not given a conclusive answer so far 
is that both systematical (hadronisation, 
higher order perturbative corrections) and statistical (4-jet decays
constitute only $\Ord10\%$ the hadronic sample)
errors spoil considerably the precision of the measurements, thus preventing
one from putting stringent bounds on $C_A$, $C_F$ and $T_R$.
However, the most serious and {\sl intrinsic} 
limitation of the analyses performed up to now is that 
they made use of energy ordering to distinguish
between quark and gluon jets and to assign the momenta 
to the final state partons\footnote{The 
two most energetic jets are identified as primary quarks. Unfortunately,
for $Q\bar Qgg$ events, in only half of the cases the two
lowest energy partons are both gluons \cite{nonAbel} !}.

A clear improvement to this approach is the one proposed in
Ref.~\cite{ioebas}. There, it was shown the superiority 
of using 4-jet samples in which two jets are tagged as $b$-jets.
In this way, one gets
a greater discrimination power between $Q\bar Qgg$ and $Q\bar Q
q\bar q$ events. First, because this way one is able to distinguish
between quarks and gluons, thus assigning the momenta 
correctly. Second, because gluon
rates are reduced by almost a factor of 2
with respect to the quark ones, such that 
differences between the two partonic components
can be more easily investigated.
Following Ref.~\cite{ioebas}, the LEP Collaborations have recently 
performed new studies \cite{done,progress}, whose preliminary
results show indeed that SUSY predictions can be more efficiently constrained. 
Furthermore, they have proved that adopting a double vertex 
tag\footnote{Which reduces considerably the statistical sample, as 
the current efficiency at LEP1 in tagging a displaced vertex is $\varepsilon
\approx30\%$ per jet.} does not 
ruin the advantages gained with particle identification.

Besides the final results of these new, improved analyses,
we want to stress that there are other
possibilities offered by the $\mu$-vertex devices, 
that can be exploited in order
to either confirm or disprove the presence of SUSY signals in 4-jet events.
This is apparent if one notices that light gluinos can also be relatively
long-lived, such that they might produce detectable
secondary vertices \cite{Cuypers}.
It is the purpose of this letter to study to
which extent
such experimental techniques can be used for detecting or ruling out SUSY
signals at
LEP1, even when no special effort is made to distinguish between
displaced vertices due to $b$-quarks and to gluinos.

The plan of the paper is as follows. In the next Section we describe
our calculations, in Section 3 we discuss the results, and
in Section 4 we summarise and conclude.

\subsection*{2. Calculation} 

In carrying out the study described here we made use of the 
{\tt FORTRAN} matrix elements 
already discussed in Ref.~\cite{MEs} and presently 
used for experimental simulations \cite{OPAL},  
upgraded with the inclusion of the gluino 
production and decay mechanisms (see also Ref.~\cite{masses})\footnote{The 
numerical values adopted for quark masses and \sm\ parameters
can be found in Ref.~\cite{MEs}.}.
The programs do not contain any approximations, the intermediate
states $\gamma^*$ and $Z$ being both inserted, and the 
masses and
polarisations of all particles in the final states
(of the two-to-four body processes) retained.
The availability of the last two options is especially important
if one considers, on the one hand, 
that in $b$-tagged samples all final states are massive,
and, on the other hand, that in proceeding to experimental fits 
one could well select restricted
regions of the differential spectra of the angular variables,
where the rates are likely to strongly depend on the spin state of the 
partons\footnote{For this reason 
we have not used the results published in literature for the gluino
decay rates, as these are averaged
over the helicities of the unstable particle.
Instead we have recomputed the relevant Feynman decay amplitudes
by preserving the gluino polarisation and by matching the latter with
the corresponding one in the production process.}.
			       
As jet finding algorithm we have adopted here the Durham (D) 
scheme \cite{schemes}. However, none of the results 
drastically depends on the choice of the jet recombination procedure and/or
the value of the jet resolution parameter, $y_{\mathrm{cut}}$.
Finally, to make clear the rest of the paper, we use the following notation:
when heavy flavour identification is implied, 
labels 1 \& 2 refer to the two tagged 
jets and 3 \& 4 to the two remaining ones.
If no vertex tagging is assumed,
jets are labelled according to their energy, 
${E_1\ge E_2\ge E_3\ge E_4}$.

\subsection*{3. Results}

\subsection*{3.1 Gluino tagging}

When dealing with tagging a secondary vertex possibly due to 
gluino decays, several points must be addressed.
First of all, one has to confine oneself to secondary vertex 
analyses only\footnote{Thus
neglecting other forms of heavy flavour tagging: such as
the high
$p_T$ lepton method, as gluinos do not decay semileptonically.}, however 
this technique 
has a larger efficiency than any other method
\cite{Squarcia}.
Second, the vertex has to be inside the detectors, so that only gluinos
with $\taugluino\Ord10^{-9}~\mathrm{s}$ can be
searched for \cite{Cuypers}. Nonetheless,
this represents an appealing opportunity,
as a substantial part of the $(\mgluino,\taugluino)$ window \cite{window}
not yet excluded by the experimental data could be covered.
The latest constraints still allow for the existence of relatively
long-lived and light gluinos, in the parameter
regions:
(i) $\mgluino\Ord1.5$ GeV and $\taugluino\Ord10^{-8}$ s;
(ii) $\mgluino\OOrd4$ GeV and $\taugluino\OOrd10^{-10}$ s
 \cite{Farrar-review,Kileng-Osland1}.

In this respect, we exploit a sort of `degeneracy' in
lifetime between $b$-quarks and gluinos, assuming that when 
making secondary vertex tagging
one naturally includes in the $2b2\mathrm{jet}$ sample also
SUSY events, in which a \gluino\ behaves as a $b$.
We call such approach `minimal trigger' procedure, as we propose 
a tagging strategy that {does not take} into account
any of the possible differences between gluinos and $b$-quarks
in 4-jet events with two secondary vertices 
(thus, in the following, we will generally speak of `vertex tagging').
There are in fact at least three obvious dissimilarities.

\noindent
(i) Their charge is different, such that one could ask that the jet
showing a displaced vertex has a null charge. This would allow one
to isolate a sample of pure SUSY events. However, we remind
the reader that measuring the charge of a low energy jet in 4-jet events
would have very low efficiency (in isolating a very broad hadronic system in
an environment with high hadronic multiplicity) and has not 
has not been attempted before.

\noindent
(ii) Gluino lifetimes much longer than $b$-lifetimes are still consistent
with experiment  (note that $\tau_b\approx
10^{-12}~\mbox{s}$), such that recognising a 
displaced vertex with decay length $d\gg3$ mm (that of the $b$) would
allow one to immediately identify gluinos. Unfortunately, most
of the $b$-tagged hadronic sample at LEP1 has been collected
via a bi-dimensional 
tagging (see, e.g., Ref.~\cite{bidimensional}). Thus,
different $d$'s could well appear the same on the event plane. Furthermore, 
tagging a $d> 3$ mm vertex would allow one
to separate gluinos with $\taugluino>\tau_b$, but this 
would not be helpful if $\taugluino\le\tau_b$.

\noindent
(iii) Other than in lifetime, $b$-quarks and gluinos can differ in mass
as well, such that one might attempt to exploit mass constraints to separate
SUSY and pure QCD events. However, on the one hand,
one could face a region of
$m_b$-\mgluino\ degeneracy 
and, on the other hand, one should cope with the ambiguities
related to the concept of mass as defined at partonic level and
as measured at hadron level.

\noindent
We emphasise that
measuring the charge of the vertex tagged jet, attempting
to disentangle different decay lengths or measuring
partonic masses could well be
further refinements of the procedure we are
proposing. These could be implemented at a later stage without
spoiling the validity of our approach.
In addition, all these aspects would necessaril need a proper
experimental analysis, which is beyond the scope of a theoretical study.

The steps of our procedure are very simple. Under the assumption that
$b$'s and $\gluino$'s are not distinguishable by vertex tagging, one
naturally retains in the 
sample all SUSY events, whereas the ordinary QCD
components are reduced by a factor of 5 ($Q\bar Qgg$) and 3 
($Q\bar Q q\bar q$).
Then, it is easy to notice that there exist
clear kinematic differences between the $Q\bar Qgg$, $Q\bar Qq\bar q$
and $Q\bar Q\gluino\gluino$ components. This is shown in Fig.~1, where
we plot the quantities $Y_{ij}=M_{ij}^2/s\equiv (p_i+p_j)^2/s$, where $ij=12$ 
or $34$ and $s=M_Z^2$. The behaviour of the curves
is dictated by the fact 
that gluinos are always secondary products in 4-jet events, 
whereas quarks and 
gluons are not (lower plots). 
When no vertex tagging is exploited and the common energy ordering is 
performed,
such differences are washed away (upper plots).
The value $m_\gluino=5$ GeV is assumed for reference, the
shape of the distributions being qualitatively the same regardless of it. 
Therefore, if one simply asks to reject events for which, e.g.,  
$Y_{12}>0.2$  and/or $Y_{34}<0.1$, one gets for the total rates of the
three components the pattern depicted in Fig.~2. 
%From the drastic
%predominance of $2Q2g$ events in the complete `unflavoured' sample 
%(no vertex tagging, top left), one first obtains that
%the total rates of ordinary QCD events are significantly reduced compared
%to those of SUSY events
%(after vertex tagging, top right),   
Notice that the drastic
predominance of $2Q2g$ events in the complete `unflavoured' sample 
has disappeared. Further,
the total rates of ordinary QCD events are significantly reduced compared
to those of SUSY events
(after vertex tagging, top right),   
and eventually the $Q\bar Q\gluino\gluino$ fraction
is always comparable to that of $Q\bar Qgg+Q\bar Qq\bar q$ events    
(when also the kinematics cut are implemented, bottom                
left). Most important, this is true {\sl independently} of the gluino mass,      
of the jet algorithm and                                                 
of the $y_{\mathrm{cut}}$ value used in the analysis.

Therefore, after our event selection, SUSY signals would certainly
be identifiable, as a clear excess in the total number of 4-jet events
with two displaced vertices. Thus,
the presence of gluinos at LEP1 could be revealed or excluded at least 
over appropriate
regions of masses and
lifetimes\footnote{And this should certainly be done
after
the appropriate MC simulations, including the details of the detectors
and of the tagging procedure
as well as a generator where \mgluino\ and \taugluino\ enter as
free parameters to be determined by a fit.}. 
We finally stress
that, as we are concerning here with total rates and not with differential 
distributions, the event number should be sufficient to render the analysis
statistically significant\footnote{In this respect we acknowledge that many
of the aspects of our approach were already employed in
Ref.~\cite{Cuypers}, however the tagging procedure sketched there 
is well beyond the statistical possibilities of the LEP1 experiments.}.
Furthermore, the ambiguities related to the fact that gluino effects
on the total number of 4-jet events are comparable in percentage to the
systematic uncertainties due to the jet hadronisation process and/or
the $y_{\mathrm{cut}}$ selection procedure 
are much less severe in ours than in the usual approach
\footnote{These are in fact the underlying difficulties 
of any analysis based on the `unflavoured' hadronic sample and/or the
jet energy ordering, which have not been overcome even in recent  
improved analyses \cite{Murayama}.}. 
However, since the key point of the present study is to exploit 
the $b/\gluino$ vertex degeneracy, a highly enriched heavy flavour sample 
should be selected in this case.

\subsection*{3.2 Gluino decays}

Before closing, we should mention that a further aspect must be
kept into account when attempting our analysis. 
It concerns the kinematics of
the gluino decays. In the most widely supported SUSY
framework \cite{Nilles}, the dominant gluino decay modes
are $\gluino\ar q\bar q\photino$ and $\gluino\ar g\photino$, where
$\photino$ represents a `photino' (better, 
the Lightest Supersymmetric Particle,
which is a superposition of the \susy\ partners
of the neutral gauge bosons of the theory).
Furthermore, the $q\bar q \photino$ channel
is, in general, largely dominant over the $g\photino$ mode 
\cite{Haber-Kane2}.

The crucial point is that in both cases the gluino decays into
a jet with missing energy. It is not our intention
to discuss the possibility
of selecting such a signature, as we are mainly concerned here with the
fact that the energy left to the hadronic system $E_h$ is above
the experimental cuts in minimal hadronic energy, which are used  
to reduce the
backgrounds (e.g.,
in Ref.~\cite{OPAL} the threshold was set equal to 3 GeV).
In Fig.~3 (first three plots) we show 
the $E_h$ spectra after the gluino decay, in both 
the channels. Two kinematic decay configurations are considered:
a massless photino, and a massive one (i.e., $\mphotino =1/2\mgluino$).
The message is that in the most likely SUSY scenario 
(i.e., three-body decay dominant and massless photino) all
gluino events should be retained in the event selection. Conversely, the
figure illustrates the percentage of these which will pass the adopted
trigger requirements.
Finally, in the bottom right plot of Fig.~3 we show 
the dependence of the SUSY rates on the value of $\mgluino$.
Below $\mgluino\approx5$ GeV, the mass suppression
is always less than a factor of 2.

\subsection*{4. Summary and conclusions}

In this paper we have stressed the importance of using at LEP1
samples of 4-jet events
in which two of the jets show a displaced vertex, and of
adopting simple invariant mass cuts based on the different kinematics of
partons in the final state. This can help to  
settle down the ongoing dispute about the existence 
of SUSY events in the data, at least for a wide range of gluino
masses and lifetimes. Those presented are theoretical results, 
which should be in the end verified by detailed MC simulations that
could even improve our event selection strategy (by exploiting differences
between $b$'s and $\gluino$'s, in charge, mass and lifetime).  
Therefore, it is our opinion that
the matter raised and procedures 
similar to the ones outlined here would deserve experimental attention. 
An enlarged and more detailed version of the present paper, which contains
a generalisation
of our results to other three jet schemes together with
a discussion of the angular variable dependence of the ordinary QCD and
SUSY rates, will be given elsewhere \cite{gluino}.
%\subsection*{Acknowledgements}
\vskip0.15cm
\noindent
This work is supported 
by the MURST, the UK PPARC, 
the Spanish CICYT project AEN 94-0936,
the EC Programme HCM, 
Network ``Physics at High Energy
Colliders'', contracts CHRX-CT93-0357 DG 12 COMA (SM) and ERBCHBICHT (RMT).
KO is grateful to the Cambridge Overseas Trust and Trinity College Cambridge 
for financial support.
%\goodbreak

\vfill
\clearpage

%%%%%%%%%%%%%%%%%%%%%%%%%%%%%%%%%%%%%%%%%%%%%%%%%%%%%%%%%%%%%%%%%%%%%%%%%%
%\end{document}
%%%%%%%%%%%%%%%%%%%%%%%%%%%%%%%%%%%%%%%%%%%%%%%%%%%%%%%%%%%%%%%%%%%%%%%%%%

\begin{figure}[p]
~\epsfig{file=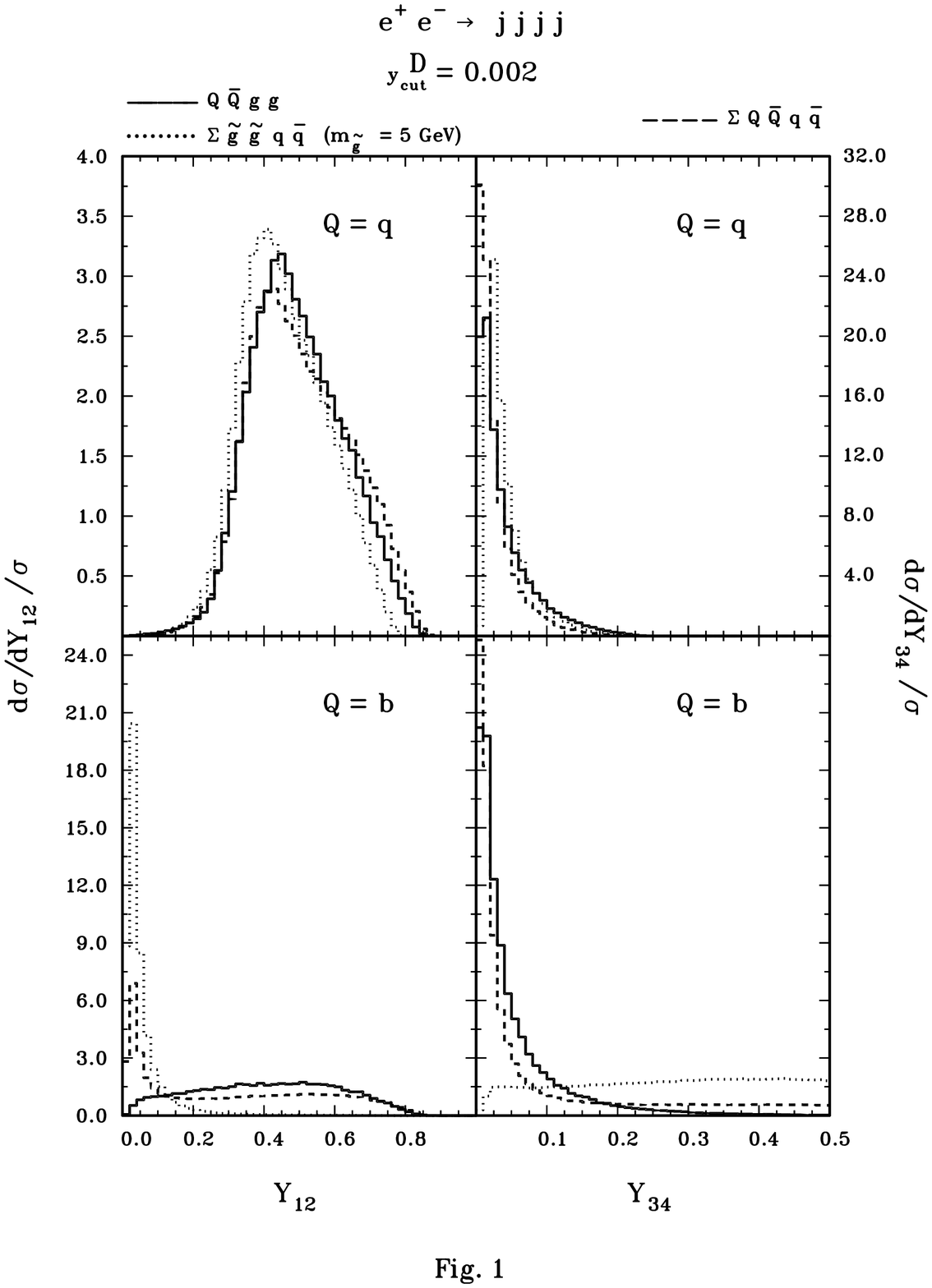,height=22cm}
\vskip0.0005cm
\noindent
{\small Distributions in the rescaled invariant masses $Y_{ij}=M^2_{ij}/s$,
where $ij=12,34$, for ordinary QCD and for SUSY 4-jet
events, in the D scheme with
$y^D_{\mathrm{cut}}=0.002$,
without ($Q=q$)  and with ($Q=b$) vertex tagging. Here, $\mgluino=5$ GeV.}
\end{figure}
\stepcounter{figure}
\vfill
\clearpage

\begin{figure}[p]
~\epsfig{file=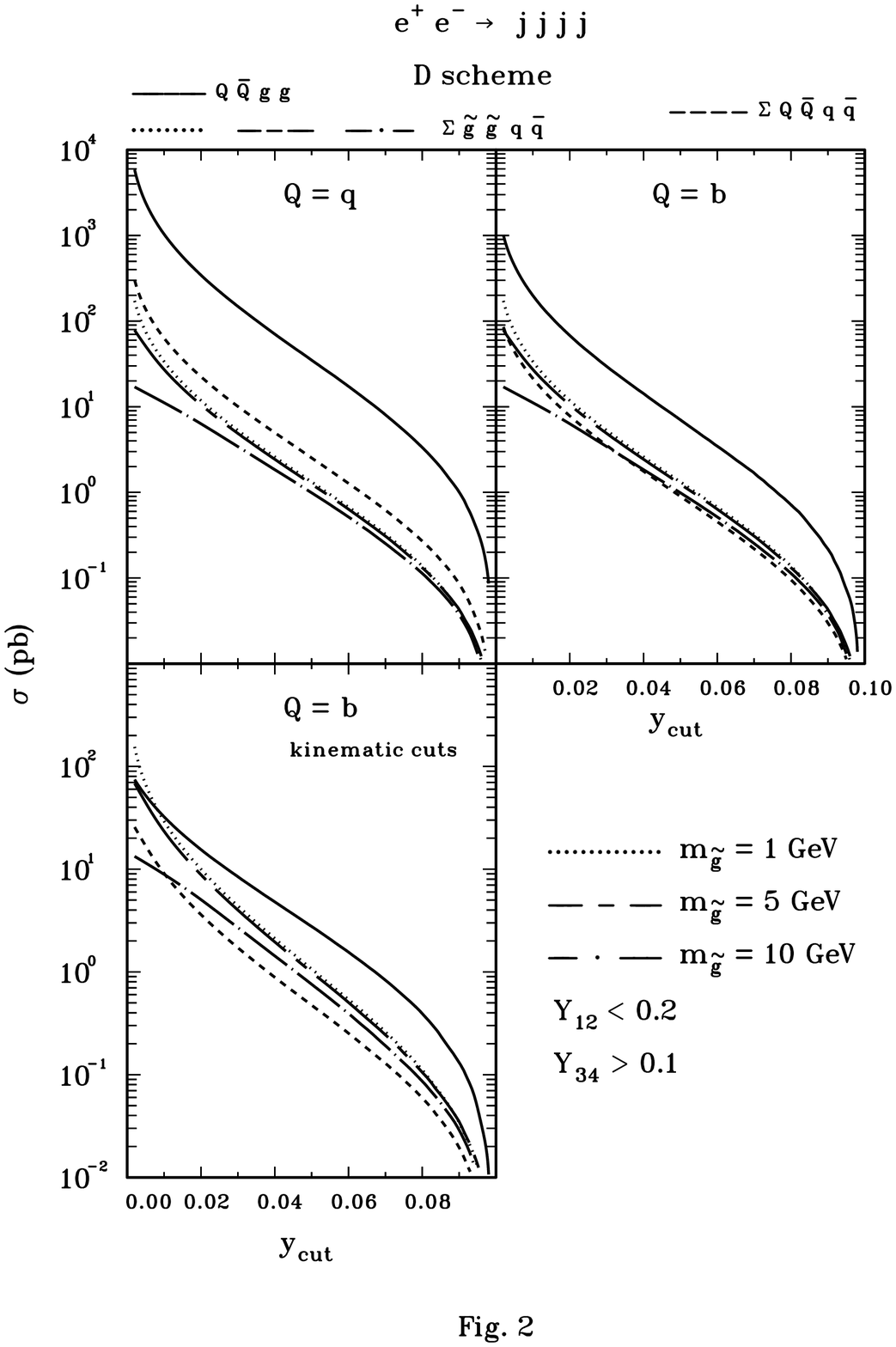,height=22cm}
\vskip0.0005cm
\noindent
{\small Total cross sections of ordinary QCD and of SUSY 4-jet 
events, in the D scheme, 
without ($Q=q$)  and with ($Q=b$) vertex tagging, 
and after the kinematic cuts, for three different values of $\mgluino$.}
\end{figure}
\stepcounter{figure}
\vfill
\clearpage

\begin{figure}[p]
~\epsfig{file=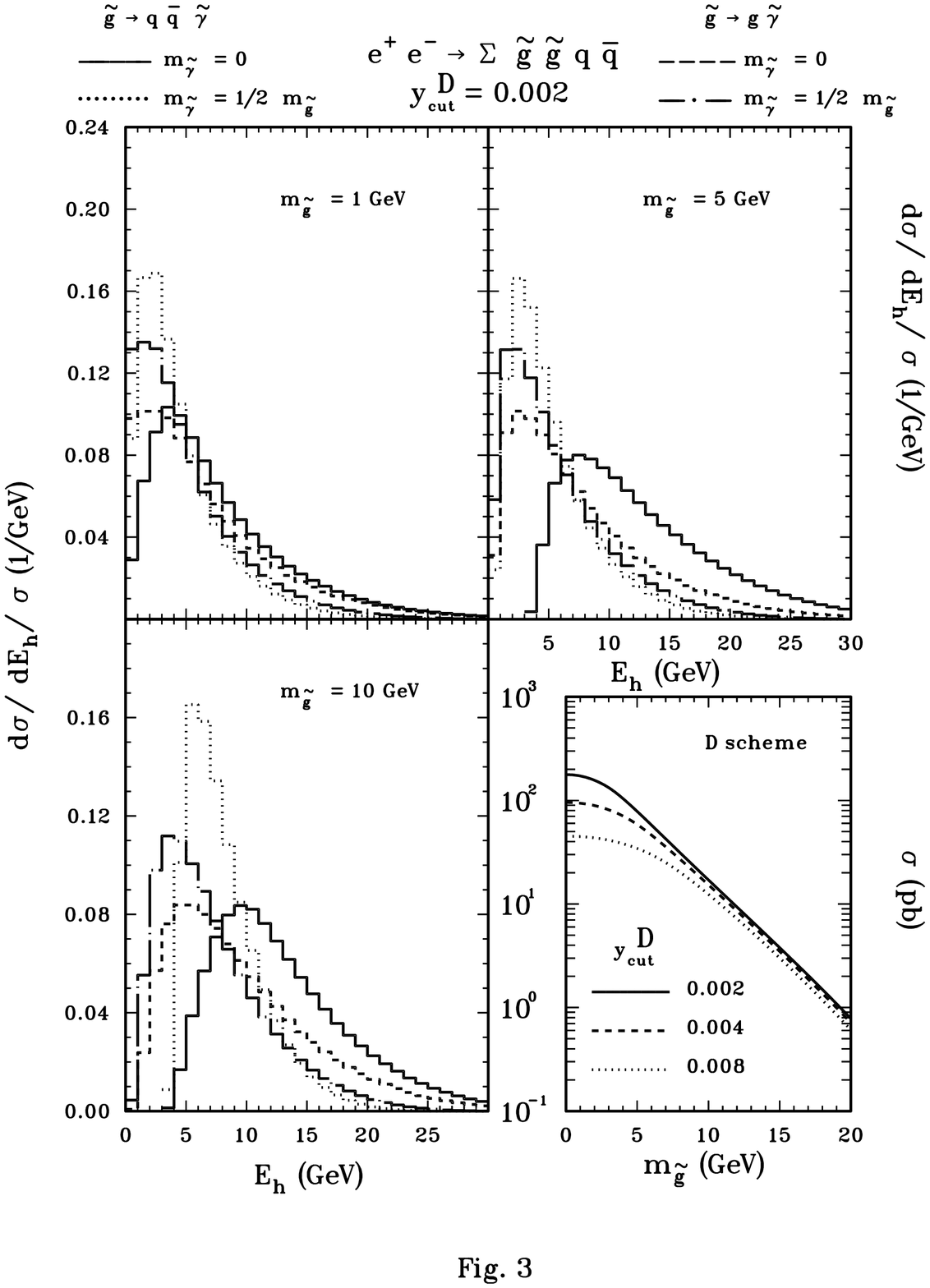,height=22cm}
\vskip0.0005cm
\noindent
{\small Differential distributions in the hadronic
energy of the `gluino jet' after the two possible SUSY decays,
in the D scheme, for various combinations of $\mgluino$ and $\mphotino$;
and total cross section of gluino events in 4-jets, as a function
of $\mgluino$ and for three different values of 
$y^D_{\mathrm{cut}}$.}
\end{figure}
\stepcounter{figure}
\vfill
\clearpage

\end{document}